\documentstyle[12pt,aaspp4,flushrt]{article}

\input{epsf}

\journalid{}{}
\articleid{}{}
\begin{document}
\def\lax    {\ifmmode{_<\atop^{\sim}}\else{${_<\atop^{\sim}}$}\fi}
\def\gax    {\ifmmode{_>\atop^{\sim}}\else{${_>\atop^{\sim}}$}\fi}
\def\gtorder{\mathrel{\raise.3ex\hbox{$>$}\mkern-14mu
             \lower0.6ex\hbox{$\sim$}}}
\def\ltorder{\mathrel{\raise.3ex\hbox{$<$}\mkern-14mu
             \lower0.6ex\hbox{$\sim$}}}
 
\long\def\***#1{{\sc #1}}
 
\title{The X-ray Nova GRS 1739-278  near the Galactic Center}

\author{K. N. Borozdin\altaffilmark{1}. M. G. Revnivtsev\altaffilmark{1}, S. P. Trudolyubov\altaffilmark{1},
N. L. Aleksandrovich\altaffilmark{1}, R. A. Sunyaev\altaffilmark{1,2}, and
G. K. Skinner\altaffilmark{3}}

\altaffiltext{1}{Space Research Institute, Russian Academy of Sciences, ul. Profsoyuznaya
84/32, Moscow, 117810 Russia}
  ¿ 
\altaffiltext{2}{Max Planck Institut fuer Asrrophysik,  KarlSchwarzschild Strasse 1, 86740
Garching hei Muenchen, Germany}

\altaffiltext{3}{University of Birmingham, Edgbaston, Birmingham, B152TT,
UK}

Recieved  Feb 5, 1998

Translated by V. Astakhov

\begin{abstract}

The soft X-ray nova GRS 1739-278 flared up in 1996 near the
Galactic center. We present the observations of this interesting source, a
black-hole candidate, with the instruments on board the Mir-Kvant module and
the RXTE satellite. The TTM data allow the spectrum of the X-ray nova to be
studied during the rise in its brightness. The source's spectrum in this
period is satisfactorily described by a power law with a gradually
increasing index. The RXTE spectra after the source passed its maximum
brightness have the appearance that is typical of the high and ultrahigh
states of black-hole candidates; they are described by a two-component
model. The broad-band spectra of the source are discussed in terms of the
processes that proceed near an accreting black hole in a binary system.  
\end{abstract} 

\section{INTRODUCTION}

Over the last decade, the X-ray observatories on board the Kvant module and
the Granat satellite have obtained spectra of several transient sources in a
wide X-ray range. A characteristic group of bright objects with a set of
common features that are combined into a class of X-ray novae stands out
from them. It should be noted that different authors put a more or less
broad sense into this notion. In this paper, we rely on the interpretation
of Sunyaev et al. (1994), in which binary systems that consist of a low-mass
optical component and a black hole belong to X-ray novae; these binaries
spend most of the time in the off state and undergo rare, extremely bright
outbursts, during which the X-ray flux from the source increases by several
orders of magnitude. The outburst light curves of such sources show a steep
rise and a prolonged quasi-exponential decline with one or more secondary
maxima. In general, the X-ray spectra of this class of objects exhibit the
same features as do the spectra of other black-hole candidates. These
include the characteristic spectral states which have been extensively
covered in the literature (see, e.g., Sunyaev et al. 1994; Tanaka and
Shibazaki 1997). The manifestation of activity in the radio range, which is
interpreted as the emission from ultrarelativistic jets and clouds of hot
plasma ejected from the immediate vicinity of the black hole, can be
considered to be a distinctive feature of many X-ray novae (see, e.g.,
Mirabel and Rodriguez 1994).

The new hard X-ray source GRS 1739-278 was discovered near the Galactic
Center on March 18, 1996 by the SIGMA gamma-ray telescope on board the
Granat satellite (Paul et al. 1996). The initial SIGMA localization of GRS
1739-278 was refined by the TTM instrument (Borozdin et al. 1996). VLA radio
observations revealed a radio source within the TTM error region (Durouchoux
et al. 1996). Mirabel et al. (1996) measured the optical/infrared flux from
this object.       

In 1996, the source was observed in the X-ray band by ROSAT (Greiner et al.
1997), Granat (Vargas et al. 1997), RXTE (Takeshima et al. 1996), and the
Kvant module of the Mir Space Station. In this paper, we focus our attention
on the results from the TTM/Kvant telescope and from the PCA and HEXTE
instruments of the RXTE satellite. Our analysis of the available data allow
us to classify GRS 1739-278 as a soft X-ray nova and a black-hole candidate.

\section{OBSERVATIONS AND DATA REDUCTION}

The Kvant module has been operating in orbit since 1987. In the past years,
numerous observations of the Galactic-center region have been carried out.
The coded-aperture TTM telescope has shown itself to be a valuable source of
information about the processes in this region of the sky. The instrument
records X-ray emission in the energy range 2 to 30 keV with an angular
resolution of about 2 arcmin in a 15ox15o field of view. The Galactic-center
region with GRS 1739-278 has repeatedly fallen within the TTM field of view
in the previous years, but a significant flux from the source was first
detected by the sum of three sessions in March 6-7, 1996. The source was
brightest during the next series of observations from February 28 through
March 5, 1996. During this series of observations, the main target was the
transient bursting pulsar GRO J1744-28 (Aleksandrovich et al. 1998). The TTM
X-ray image of a sky area near the Galactic center on March 1, 1996 is shown
in Fig. 1. The relative confidence of detecting the brightest sources during
these observations is along the vertical axis. The TTM coded aperture makes
it possible to reliably separate the fluxes from various sources in the
densely populated region of the Galactic center. However, the presence of
the very bright source GRO J1744-28 at the center of the field of view
raises considerably the background level when other sources are recorded and
reduces their relative contribution to the limited telemetry flow, resulting
in an increase in the TTM dead time and in a decrease in the effective
observing time. In order to obtain statistically significant spectra, we
added the data for successive sessions, if the total flux from the source
did not change too much from session to session. Interestingly, during the
observations of GRO J1744-28 in February-March 1996, GRS 1739-278 also fell
within the field of view of the HEXE spectrometer on board the Kvant module,
which is sensitive to the hard X-ray energy band. The HEXE instrument is
incapable of producing images and cannot separate the fluxes from GRO
J1744-28 and GRS 1739-278, which are in close proximity to each other.
Nevertheless, the shape of the HEXE spectrum during these observations
(Borkus et al. 1997) leads us to conclude that the flux from GRS 1739-278 at
hard energies < 100 keV is significant.   

The observations in February-March 1996 and the TTM fluxes from the source
are given in Table 1. By the next series of Kvant observations of this sky
region in October 1996, the flux from the source had dropped below the TTM
detection threshold. An analysis of the Kvant archival data shows that no
significant flux has ever been detected by TTM from GRS 1739-278, except for
the observations during the source's outburst in February-March 1996. In
this case, the lower detection limit in the range 2-30 keV is 0.001 of the
flux from the Crab nebula (1 mCrab), as estimated from the sum of all
sessions.

The RXTE satellite observed the X-ray nova GRS 1739-278 on March 31, 1996
and nine more times from May 10 through May 29 of this year, each with an
exposure of several kiloseconds. These observations are given in Table 2.
The total exposure was about 24 ks.   

The RXTE satellite have two aligned spectrometers with a 1o.0 field of view
each: a set of five xenon proportional counters PCU-PCA with a maximum
sensitivity in the energy range 4-20 keV and the HEXTE scintillation
spectrometer which consists of eight Na I(Tl)/Cs I detectors sensitive to
the range 15-250 keV. The HEXTE detectors are combined into two clusters of
four detectors each; each cluster observes the source in turn. Concurrently,
the second cluster measures the instrumental and external X-ray background.

The large collecting area of the PCA detectors makes it possible to
considerably reduce the statistical errors when constructing the spectra of
bright X-ray sources. The greatest contribution to the measurement errors of
the flux in individual spectral channels comes from the systematic errors:
the fitting errors when constructing the response matrix in the soft energy
band and the errors in computing the background at harder energies. We used
the latest versions of the FTOOLS software package, which was developed by
the RXTE team, to reduce the PCA data [see Jahoda et al. (1997) for
computations of the matrix; see Stark et al. (1997) for simulations of the
background.

It should be noted that the general normalization of the PCA spectra differs
from the normalization of other instruments. For example, the general
normalization of the PCA spectrum of the Crab nebula differs from the
ASCA/GIS normalization by 10 to 20\% (see, e.g., Fukazawa et al. 1996).    

Note also that the PCA spectrum of the Crab nebula exhibits noticeable (of
the order of several percent) deviations from a power law for energies near
5 and 10 keV, which appear to be a result of inaccuracies in fitting the
edges of xenon absorption. In order to compensate for the PCA systematic
errors when reconstructing the spectra, we included an additional component
of 2\% in the measurement error of the flux in each individual energy
channel. A similar analysis performed for the HEXTE spectra of the Crab
nebula revealed no significant systematic errors introduced by the response
matrix or by the method of background subtraction up to energies at which
the count rate from the source was less than 1\% of the background count
rate (see also Rothschild et al. 1997). In the HEXTE case, the errors were
calculated by the standard method, without introducing an additional
systematic error.

\section{LOCALIZATION}

GRS 1739-278 was first localized by the SIGMA telescope of the Granat
satellite (Paul et al. 1996). Shortly afterward, the position of the source
was refined by using the TTM data: R.A. = 17h42m40s, Decl. = --27o45'.8
(epoch 2000, the uncertainty is 1 arcmin). These coordinates, published in
an IAU Circular (Borozdin et al. 1996), were used to plan the subsequent
observations in the X-ray (Takeshima et al. 1996; Dennerl and Greiner 1996),
radio (Durouchoux et al. 1996), and optical (Mirabel et al. 1996) ranges,
which made it possible to identify GRS 1739-278 with an optical object and
to record a radio flux from the source. The SIGMA, TTM, and ROSAT error
regions of GRS 1739-278 and the positions of the optical and radio sources
are shown in Fig. 2. 

\section{THE LIGHT CURVE OF THE SOURCE}

Figure 3 shows the 2-10-keV light curve of the source in 1996, as
constructed from the data of the All-Sky Monitor (the ASM instrument) on
board the RXTE satellite and of the TTM telescope on board the Kvant module.
Interestingly, the flux from the source was recorded by TTM as early as on
February 6-7, 1996, i.e., more than a month before the presumed maximum in
the range 2-10 keV, which had occurred approximately by March 10. The ROSAT
observations of the source revealed a bright halo, which allowed Greiner et
al. (1997) to assume that a transition of the source to the high state began
as early as in November-December 1995. Such a protracted rise in the flux is
not typical of X-ray novae, which normally reach their maximum light within
a few days after the onset of an outburst. According to the TTM and ASM
data, the flux from the source rose not monotonically, but underwent
irregular variability. The local maximum recorded by TTM on March 1 stands
out against this background. Within the first month after the bright
outburst with a flux at maximum light close to the flux from the Crab
nebula, a quasi-exponential decline with a characteristic time of about 34
days was observed, and the flux from the source was then nearly constant for
two weeks (from the 30th through the 43rd day after the maximum);
subsequently, the brightness decline of the source continued, but was now
slower with a characteristic time of about 48 days (from the 44th through
the 68th day after the maximum). This temporal behavior is typical of many
observed X-ray novae. In 1996, three more outbursts were observed from GRS
1739-278. Note the observed tendency for the rise time of the flux to
increase and for the decay time to decrease. Finally, the flux from the
source dropped below the sensitivity threshold of the RXTE All-Sky Monitor
by October-November 1996. During the series of observations of the
Galactic-center region in October 1996, the source was not detected by the
TTM telescope either.  

\section{THE SPECTRUM}

The TTM data were obtained on the ascent of the source's light curve to a
maximum in the range 2-10 keV. These data are of particular interest,
because spectroscopy of X-ray novae is generally performed on the descent of
the light curve. The spectrum for all days of TTM observations is
satisfactorily fitted by a power law with absorption. Unfortunately, the
insufficient statistical significance of the results prevents an in-depth
analysis of the parameter evolution in individual sessions. However, after
adding up the data for several successive sessions, we can trace the general
long-term trends in the changes of the spectral shape. 

The parameters of the power-law fit to the spectrum of GRS 1739-278 for
different days of observation are given in Table 1. As with many other
sources in the Galactic-center region, the source's spectrum in the soft
band is characterized by a substantial absorbing column density. Greiner et
al. (1997) estimated the absorbing column density to be 2.01022 cm--2;
different estimation methods yield values between 1.61022 and 2.61022 cm--2.
The power-law fit to the combined TTM spectrum (see Table 2) adequately
describes the data for NHL = $(2.79\pm0.52)10^{22}$ cm$^{-2}$. When fitting the TTM
spectra for individual days, we fixed the absorbing column density in order
to trace the evolution of the spectral hardness by the change in the slope
of the power-law fit. Our results for two values of NHL --- one was taken
from Greiner et al. (1997), and the other was obtained by fitting the
combined TTM spectrum for March 1-5, 1996 --- are given in Table 1.

With the exact slope of the power-law spectrum depending on the assumed
absorbing column density, the main tendency for the spectral hardness to
decrease with time is preserved (see Table 1 and Fig. 4). Interestingly,
this tendency also shows up for the sessions of March 3-5, although the flux
from the source on these days was lower than its flux on March 1. A similar
behavior of the spectrum at the initial phase of outburst has been
previously noted in other X-ray novae, in particular, in Nova Muscae 1991
(Lapshov et al. 1992; Ebisawa et al. 1994) and KS 1730-312 (Borozdin et al.
1995; Trudolyubov et al. 1996).

The combined spectrum, as constructed from the observations on March 1-5,
1996, is also satisfactorily described by a power law with absorption (see
Table 3). However, such a spectrum with a photon index [alpha] > 2.0 is not
typical of X-ray novae and other black-hole candidates (for a discussion,
see below). Since the subsequent PCA/RXTE observations revealed a bright
soft component in the spectrum of GRS 1739-278 (see Table 4), we surmised
that a similar component must also be present in the TTM spectrum, together
with the hard power-law component that is typical of black-hole candidates
in the low and off states. In our fitting, we fixed the absorbing column
density at the same values as in the case of fitting individual sessions by
a simple power law. In addition, we fixed the slope of the power-law
component at [alpha] = 1.5, 1.7, and 2.0, i.e., we specified the values in
the range that is typical of the low state for black-hole candidates (recall
that the spectrum of this hardness was recorded during the first TTM
detection of GRS 1739-278). The fits in Table 3 show that the spectrum can
be satisfactorily described by a two-component model at reasonable fitting
parameters (see Table 4 for a comparison). At the same time, it is clear
that the TTM data do not allow sufficiently stringent constraints to be
placed on the model parameters.        
  
The RXTE spectra have the appearance that is typical of the spectra of X-ray
novae (see, e.g., Grebenev et al. 1991; Sunyaev et al. 1988, 1994). In
general, such spectra are well fitted by a "multicolor" accretion disk model
(Makishima et al. 1986) in the soft part of the spectrum and by a power-law
component at high energies. We also used this model to fit the RXTE spectra
of GRS 1739-278, although the standard model of an optically thick accretion
disk is not self-consistent for the parameters under consideration (see
Shakura and Sunyaev 1973). 

During the observations on March 31, 1996, the source was in the ultrahigh
spectral state with a bright soft component and a hard power-law tail (see
Figs. 5 and 6). The parameters of the fitting of the source's spectra by a
model that includes the emission from an optically thick, geometrically thin
accretion disk and a hard power-law component are given in Table 4. The
combined PCA and HEXTE data allow us to trace the source's spectrum up to
energies 200 keV. During these observations, the power-law component in the
spectrum was particularly strong. The RXTE data show no exponential cutoff
exp(--E/Ef) in the spectrum at high energies. The lower limit on Ef, as
estimated from the combined PCA (15-30 keV) and HEXTE (15-200 keV) data, is
150 keV (at the 68\% confidence) or 93 keV (at the 94\% confidence). 

During the subsequent observations in May 1996, the power-law component
considerably weakened, and the source ceased to be recorded by HEXTE.
Subsequently, all parameters of the hard power-law component were determined
from the PCA data. In Fig. 7, the main parameters of our fit are plotted
against time.

As was already noted above, at the existing brightness of the source at
energies above 15-20 keV, the uncertainties in the spectrum are mainly
attributable to systematic errors in the background subtraction.
Accordingly, even when a hard tail was reliably detected in the spectrum of
GRS 1739-278, we could not accurately determine the photon index of the hard
power-law component and, in some cases, fixed its value at [alpha] = 2.4.
When determining the 15-20-keV flux, we also added a systematic error of
$3\times10^{-12}$ erg/s/cm$^2$, which accounts for 3-5\% of the background flux in
this band. The contribution (in percents) of the source's flux in the band
15-20 keV (which is dominated by the hard component) to the total 2-20-keV
flux is plotted against time in Fig. 7.      

\section{DISCUSSION}

Although X-ray novae share several common features, which allow them to be
singled out as a special class of astrophysical objects, individual
representatives of this class do not normally exhibit a complete set of
characteristic properties. For GRS 1739-278, we have already noted an
unusually slow rise in the flux before the maximum of the light curve.
However, a similar behavior at the beginning of outbursts was also observed
in other transients --- black-hole candidates --- A 1524-62, GX 339-4, and
GRS 1915+105 (Kaluzienski et al. 1975; Harmon et al. 1994; Sazonov et al.
1994). Within the first weeks after the maximum, the light curve of GRS
1739-278 was typical of X-ray novae --- a quasi-exponential decline with a
secondary maximum. The characteristic time of the flux decline is close to
the mean for the light curves of X-ray transient sources of different nature
(see Wan Chen et al. 1997). There are several models to explain the nature
of the quasi-exponential flux decline (see, e.g., Lyubarskii and Shakura
1987; King and Ritter 1997). For instance, King and Ritter (1997) proposed
to explain the characteristic flux decline in X-ray novae in terms of the
model of the development of instability in the accretion disk maintained by
hard X-ray radiation from the inner disk regions. In this case, the disk can
return to its initial cold state only after the bulk of its mass is accreted
onto the compact object. The secondary maximum in the light curve is
accounted for by the development of thermal instability in the outer disk
regions under the effect of a bright X-ray outburst. If the cold dense disk
is not all exposed to fairly intense radiation during the first outburst,
then additional maxima can appear in the light curve. In the case of GRS
1739-278, five such maxima can be noted before the flux drops to 20 mCrab.
The interval between maxima (30-50 days for GRS 1739-278) is determined by
the time it takes for a perturbation to be transmitted in the disk through
viscosity or through the heating wave that propagates in the disk. This
model not only satisfactorily describes qualitatively the light curve in
Fig. 2, but also gives reasonable order-of-magnitude estimates of the disk
parameters, such as the outer radius ($10^11$ cm) and the kinematic viscosity
($10^{15}$ cm$^2$/s). At the same time, this model makes no assumptions about the
nature of the central X-ray source, i.e., it cannot be used to distinguish
systems with a black hole and with a neutron star as the binary's compact
component.

Spectroscopy of black-hole candidates over a wide energy range has revealed
several spectral states typical of these sources. In the low state, the
spectrum is very hard and can be fitted by a power law with an index of
1.5-2.0 up to energies 100 keV or higher. In the high state, the spectrum is
dominated by a bright soft component, which is described by the spectrum of
a "multicolor" accretion disk with a characteristic temperature of 0.7-1.5
keV; the flux at energies above 10 keV is low and is often undetectable. Of
particular interest is the ultrahigh state; its spectrum exhibits a
noticeable power-law component with a characteristic index 2.5, in addition
to a bright soft component similar to the high-state spectrum. Since the
ultrahigh state has not been observed so far in systems whose compact
component is a neutron star, this type of spectrum can be considered as
evidence for the presence of a black hole in the system.

The RXTE spectrum of the source on March 31, 1996 is a typical
ultrahigh-state spectrum of black-hole candidates, which was observed in
many soft X-ray novae. According to the Kvant and Granat observations, GS
2000+25 (Sunyaev et al. 1988), GRS 1009-45 (Sunyaev et al. 1994), GRS
1124-684 (Gil'fanov et al. 1991; Grebenev et al. 1992), and KS 1730-312
(Borozdin et al. 1995; Trudolyubov et al. 1996) had similar spectra. In
1997, the same type of spectrum was observed from the X-ray nova XTE
J1755-324 (Revnivtsev et al. 1998). The high statistical significance of the
RXTE data allow us to determine the fitting parameters with a high accuracy
and to study their variations with the total flux from the source. Figure 8
shows a plot of Rin against the 3-25-keV flux. There is a clear tendency for
this parameter to increase with decreasing flux. This result is obviously
inconsistent with the simple (and incorrect---see, e.g., Grebenev et al.
1991) interpretation of Rin as the inner disk radius corresponding to the
radius of the last stable orbit of the accretion disk which is encountered
in the literature. As was already noted above (Grebenev et al. 1995;
Trudolyubov et al. 1996), Rin has a more complex physical meaning. Figure 8
shows that the correlation between the fitting parameters of the soft
spectral component differs from Rin  Tin--4/3 predicted by the standard
accretion-disk theory (Shakura and Sunyaev 1973) for the blackbody zone of
the accretion disk. At the same time, the detected flux dependences of the
disk parameters provide evidence for the models that attribute the change in
the spectral state of black-hole candidates to a change in the
characteristic radius of the blackbody zone in the accretion disk (see,
e.g., Ebisawa et al. 1996; Esin et al. 1997).              

Of particular interest is the detection of a weak power-law component in the
observed RXTE spectrum of the source in May 1996. For the instrument with a
lower sensitivity, this spectrum is fully described by the spectrum of a
"multicolor" disk. This state is commonly referred to as high, in contrast
to the ultrahigh state with a two-component spectrum. The RXTE observations
reveal a weak power-law component in the high-state spectrum of black-hole
candidates; in this case, we can thus also speak of a two-component
spectrum.       

Of special interest in interpreting the ultrahigh-state spectra are the
models that explicitly assume the absence of an emitting surface of the
compact component in the binary system. For instance, it is interesting to
consider the model in which an optically thick, geometrically thin disk
described by the standard model (Shakura and Sunyaev 1973) lies outside,
while in the inner region the emitted soft X-ray photons are Comptonized by
relativistic electrons in an advective flow of accreted matter. An
analytical treatment and numerical simulations of such a system make it
possible to obtain a good fit to the two-component spectrum of black-hole
candidates (Titarchuk et al. 1998). This type of spectrum was recorded by
the RXTE instruments from GRS 1739-278; the absence of an exponential cutoff
at high energies suggests that the hard component resulted from the
Comptonization of soft X-ray emission by a rapid convergent flow with the
exponential cutoff presumably at $E\sim m_ec^2$ (Titarchuk et al. 1997), rather
than by hot electrons of the corona or of some other region (with the
exponential cutoff at $E\sim3kT_e$  100-200 keV). At the same time, because of
the statistical errors of the HEXTE spectrum at energies above 100-150 keV,
the possibility of the cutoff in the spectrum at energies 150-200 keV cannot
be ruled out.              

The TTM spectrum taken in March 1996 during the rise in the flux from the
source differs from any of the set of typical spectra for black-hole
candidates. It is well described by a power law with absorption and does not
require the introduction of an additional soft blackbody component. At the
same time, the slope of the power-law component (2.3-2.7) is much steeper
than the typical value for the low state of black-hole candidates (1.5-2.0).
A similar spectrum was observed by TTM and SIGMA from KS 1730-312 (Borozdin
et al. 1995; Trudolyubov et al. 1996). At that time, a hard power-law
spectrum was recorded on the first day, a spectrum with a slope of 2.7 two
days later, and a spectrum typical of the ultrahigh state of black-hole
candidates another day later. Thus, in this case, the slope of the power-law
fit to the spectrum also increases with flux. Another example of such a
change in the spectrum was observed by the GRANAT and GINGA satellites
(Lapshov et al. 1992; Ebisawa et al. 1994): during the rise in the flux from
GRS 1124-682 (the X-ray nova Muscae 1991) between January 10 and 14, 1991,
the slope of the power-law component increased from 2.24 to 2.62, with the
soft component of the spectrum being much weaker at that time then that
during the subsequent observations of the ultrahigh state. The above
examples show that the power-law shape of the spectrum with a variable slope
is typical of soft X-ray novae during their flux rise before the primary
maximum. There is a tendency for the spectrum to steepen as one approaches
the peak flux. A similar tendency was observed by the Granat instruments
during the outburst of the black-hole candidate GX 339-4 in 1991 (Grebenev
et al. 1991, 1993).                             

In all these cases, we appear to see the formation of a geometrically thin,
optically thick accretion disk in a hot, optically thin cloud. Initially,
the number of photons from the disk is small, and their Comptonization by
hot electrons of the optically thin medium dominates; the spectrum typical
of the low state of black-hole candidates is formed (Sunyaev and Truemper
1979; Sunyaev and Titarchuk 1980). As the optically thick disk is formed,
the number of soft photons increases, and, accordingly, the total flux from
the source rises, while the hardness of the recorded spectrum decreases. If
we represent the source's spectrum early in March 1996 as the sum of two
components, then the energy fluxes for each component in the 2-20-keV band
are comparable in order of magnitude (see Table 3), in sharp contrast to
their ratios for the ultrahigh, let alone high states. Interestingly, the
total 2-30-keV flux recorded by TTM early in March is approximately equal to
or even greater than the PCA flux for the ultrahigh state. Thus, in this
case, the spectral state of the source is not a function of the object's
luminosity alone, but is determined by the dynamics of the processes in the
system.               

\section{CONCLUSION}

The X-ray source GRS 1739-278 have several properties that allow it to be
reliably classified as a black-hole candidate and a soft X-ray nova. The
light curve after the maximum exhibits a quasi-exponential decline with
secondary maxima. The source's spectrum at that time corresponds to the
ultrahigh- and high-state spectra of black-hole candidates typical of soft
X-ray novae. The fact that the object belongs to this class is confirmed by
optical and radio observations. A characteristic difference of this source
from other objects of the same class is a protracted rise in the flux. 

The RXTE data reveal variations of the fitting parameters for the soft spectral component with X-ray flux. This result is inconsistent with the simple interpretation of Rin as the radius of the last stable orbit of the accretion disk, but, at the same time, provides evidence for the models that attribute the change in the spectral state of black-hole candidates to a change in the characteristic radius of the blackbody zone in the accretion disk.

The high sensitivity of the RXTE instruments made it possible to detect a
weak power-law component in the high-state spectrum of the black-hole
candidate. Thus, in this case, we can also speak of a two-component spectrum
similar to the ultrahigh-state spectrum, but with a much weaker power-law
component.   
  
During the TTM observations, a power-law spectrum of variable hardness was
recorded before the primary maximum, which shows a clear tendency to steepen
as one approaches the primary maximum. The total recorded flux in this state
was no lower than the flux during the RXTE observations of the ultrahigh
state. Similar spectra have also been previously obtained during
observations of the novae KS 1730-312 and GRS 1124-682; we can thus speak of
a clear pattern in the behavior of soft X-ray novae during the flux rise,
which is obviously attributable to the formation of an optically thick
accretion disk in the system, or a considerable decrease in the inner disk
radius.        

\section{ACKNOWLEDGMENTS}

This study was supported in part by the Russian Foundation for Basic
Research (96-02-18544 and 96-15-96343) and the INTAS (93-3364-ext). One of
the authors (K.B.) wishes to thank L.G. Titarchuk for a discussion of the
results, S.A. Grebenev and S.Yu. Sazonov for valuable remarks, and A.N.
Ananenkova, S.V. Lavrov, and A.N. Emel'yanov for help in the preliminary
reduction of the TTM data. The RXTE data were extracted from the HEASARC
electronic archive operated by the Goddard Space Flight Center (NASA, USA).

\clearpage

\begin{deluxetable}{cccccc}

\small
\tablecaption{TTM/Kvant observations of GRS 1739-278 during its outburst in
1996. The slopes of the power-law fit to the spectrum ($\alpha$) are given
for two absorbing column densities (NHL)}
\tablecolumns{6}

\tablehead{
\colhead{Date}&
\colhead{Exposure,}&
\colhead{$N_{HL}$,}&
\colhead{$\alpha$}&
\colhead{$\chi^2_{27}$}&
\colhead{Flux (2-20 keV),}\nl
\colhead{}&
\colhead{s}&
\colhead{$\times10^{22}, cm^{-2}$}&
\colhead{}&
\colhead{}&
\colhead{$\times10^{-9}$, erg/cm$^2$/s}
}
\startdata
6-7/02/96 & 2749 & 2.0 & $1.52\pm0.32$ & 0.72 & $1.9\pm0.4$\nl
          &      & 2.8 & $1.59\pm0.34$ & 0.71 &     \nl
 28/02/96 &  882 & 2.0 & $2.09\pm0.18$ & 0.48 & $4.2\pm0.5$\nl
          &      & 2.8 & $2.18\pm0.19$ & 0.46 &     \nl
  1/03/96 & 1650 & 2.0 & $2.36\pm0.07$ & 1.28 & $11.8\pm0.5$\nl
          &      & 2.8 & $2.48\pm0.08$ & 1.07 &     \nl
3-4/03/96 & 2030 & 2.0 & $2.44\pm0.09$ & 0.59 & $7.6\pm0.5$\nl
          &      & 2.8 & $2.55\pm0.10$ & 0.65 &     \nl
  5/03/96 & 1647 & 2.0 & $2.58\pm0.08$ & 1.05 & $9.9\pm0.5$\nl
          &      & 2.8 & $2.72\pm0.09$ & 0.95 &     \nl
\enddata

\end{deluxetable}

\clearpage

\begin{deluxetable}{cccccc}

\small

\tablecolumns{5}

\tablecaption{RXTE observations of GRS 1739-278 in March-May 1996}

\tablehead{
\colhead{\#} &
\colhead{Date} &
\colhead{Time, UT}&
\multicolumn{2}{c}{Effective exp., s}\\
\cline{4-5}\\
&
&
&
\colhead{PCA$^a$}&
\colhead{HEXTE$^b$}
}
\startdata
1&31/03/96&18:08:48 - 20:50:56&5215&2066\\
2&10/05/96&07:08:00 - 07:56:00&1365&\\	
3&11/05/96&18:58:56 - 20:05:52&2421&\\
4&12/05/96&03:23:28 - 03:45:52&1117&\\
5&13/05/96&17:05:20 - 17:50:56&2110&\\
6&14/05/96&13:43:28 - 14:48:00&2886&\\
7&15/05/96&17:04:32 - 18:25:52&325&\\
8&16/05/96&23:48:32 - 24:34:56&1691&\\
9&17/05/96&12:24:32 - 13:26:56&3398&\\
10&29/05/96&02:47:28 - 03:57:52&2529&\\
\enddata

\tablenotetext{a}{dead time corrected}
\tablenotetext{b}{for each cluster}
\end{deluxetable}

\clearpage

\begin{deluxetable}{cccccc}
\small

\tablecaption{Parameters of the power-law and two-component fits to the
combined TTM/Kvant spectrum of GRS 1739-278 on March 1-5, 1996}
\tablecolumns{6}

\tablehead{
\multicolumn{5}{c}{Parameters}&
\colhead{$\chi^2$(dof)}\\
\cline{1-5}\\
\colhead{$T_{in}$, keV}&
\colhead{$R_{in}\sqrt(cos(\theta))$, $km^a$}&
\colhead{$N_{HL}\times10^{22}cm^{-2}$}&
\colhead{$\alpha$}&
\colhead{$\%^b$}&
\colhead{}
}
\startdata
\cutinhead{Power law}
 & & $2.79\pm0.52$ & $2.58\pm0.10$ && 18.5(26)\nl
\cutinhead{``multicolor'' blackbody disk+power law}
 $1.10\pm0.08$ & $20.7\pm2.8$ & 2.8 (fixed) & 1.5(fixed.) & 45& 32.3(26)\nl
 $1.05\pm0.07$ & $22.3\pm3.5$ & 2.8 (fixed) & 1.7 (fixed)& 38&27.4(26)\nl 
 $0.96\pm0.08$ & $24.2\pm5.1$ & 2.8 (fuxed) & 2.0 (fixed)& 27& 21.2(26)\nl
 $1.23\pm0.08$ & $15.5\pm2.0$ & 2.0 (fixed) & 1.5 (fixed)& 46& 27.0(26)\nl
 $1.17\pm0.08$ & $15.9\pm2.4$ & 2.0 (fixed) & 1.7 (fixed)& 40& 23.6(26)\nl
 $1.13\pm0.10$ & $15.3\pm3.1$ & 2.0 (fixed) & 2.0 (fixed)& 28& 19.7(26)\nl
\hline
\enddata

\tablenotetext{a}{assuming 8.5 kpc distance}
\tablenotetext{b}{The contribution of the soft component to the total flux
in 2--20 keV energy band}

\end{deluxetable}

\clearpage

\begin{deluxetable}{ccccccc}

\small

\tablecolumns{7}

\tablecaption{Parameters of the fitting of the PCA and HEXTE/RXTE spectra of GRS
1739-278 by a two-component model with a soft "multicolor" disk component
and a hard power-law component}

\tablehead{
\colhead{\#} &
\colhead{$T_{in}$,keV} &
\colhead{$R_{in}\sqrt{cos(\theta)}$, km$^a$}&
\colhead{$\alpha$}&
\colhead{$F(3-25 keV)^b$}&
\colhead{$F_{pl.}^b$}&
\colhead{$\chi^2(dof)$}
}
\startdata
1&$1.208\pm0.007$&$13.6\pm0.2$&$2.31\pm0.04$&$5.41\pm0.01$&$1.33\pm0.04$&106.8(51)\\
2$^Ó$&$1.021\pm0.010$&$19.7\pm0.7$&2.4(fixed)&$3.46\pm0.01$&$0.11\pm0.01$&30.8(40)\\
3$^Ó$&$0.999\pm0.008$&$21.2\pm0.7$&2.4(fixed)&$3.53\pm0.01$&$0.12\pm0.01$&57.6(40)\\
4$^Ó$&$0.996\pm0.009$&$21.3\pm0.8$&2.4(fixed)&$3.51\pm0.01$&$0.14\pm0.01$&49.1(40)\\
5$^Ó$&$0.989\pm0.009$&$21.3\pm0.7$&2.4(fixed)&$3.36\pm0.01$&$0.13\pm0.01$&51.5(40)\\
6$^Ó$&$0.989\pm0.009$&$21.3\pm0.8$&2.4(fixed)&$3.40\pm0.01$&$0.16\pm0.01$&47.3(40)\\
7$^Ó$&$0.998\pm0.010$&$20.5\pm0.8$&2.4(fixed)&$3.31\pm0.01$&$0.12\pm0.01$&41.4(40)\\
8$^Ó$&$0.983\pm0.009$&$21.5\pm0.8$&2.4(fixed)&$3.35\pm0.01$&$0.16\pm0.01$&53.8(40)\\
9$^Ó$&$0.984\pm0.008$&$21.1\pm0.7$&2.4(fixed)&$3.23\pm0.01$&$0.14\pm0.01$&62.6(40)\\
10$^Ó$&$0.984\pm0.009$&$22.4\pm0.8$&2.4(fixed)&$3.72\pm0.01$&$0.24\pm0.01$&41.6(40)\\
\enddata
\tablenotetext{a}{ assuming 8.5 kpc distance}
\tablenotetext{b}{ in 10$^{-9}$erg/s/cm$^2$}
\tablenotetext{Ó}{In our reduction of these data,
we set the systematic error equal to 5\% in each energy channel in order to
increase the relative confidence of the points of the hard tail for the
power-law fit}

\end{deluxetable}

\begin{figure}
\epsfxsize=17cm
\epsffile{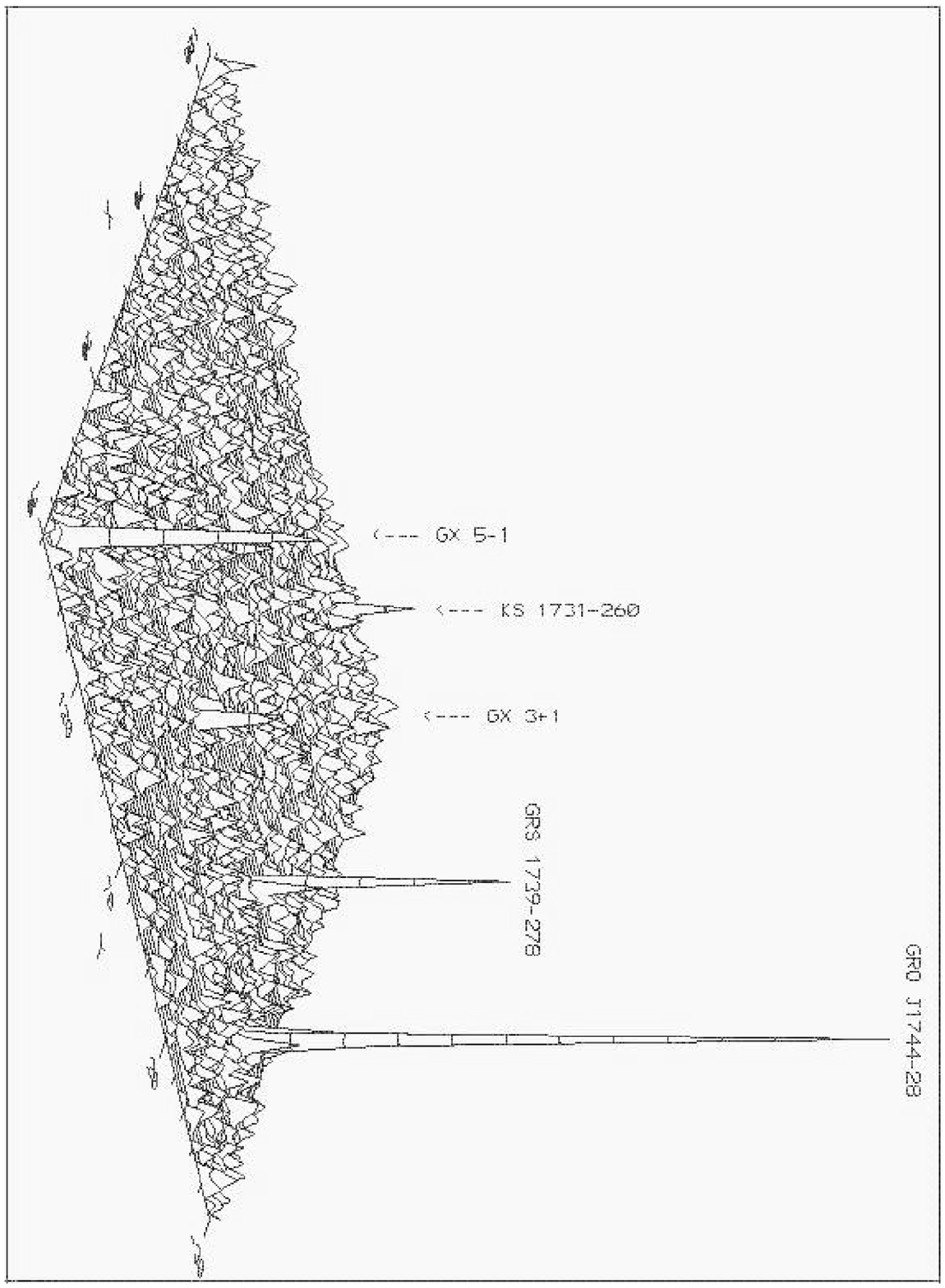} 

\caption{Part of the X-ray image of the Galactic-center region in the
2-30-keV band obtained by the TTM telescope on board the Mir-Kvant module on
March 1, 1996. The positions of the brightest (at the time of observation)
sources in this region of the sky are marked. The coordinates along the X
and Y axes are given in TTM pixels (1 pixel = 1.86 arcmin). The vertical
sizes of the peaks correspond to the confidence of detecting individual
sources. }
\end{figure}
     
\begin{figure}
\epsfxsize=17cm
\epsffile{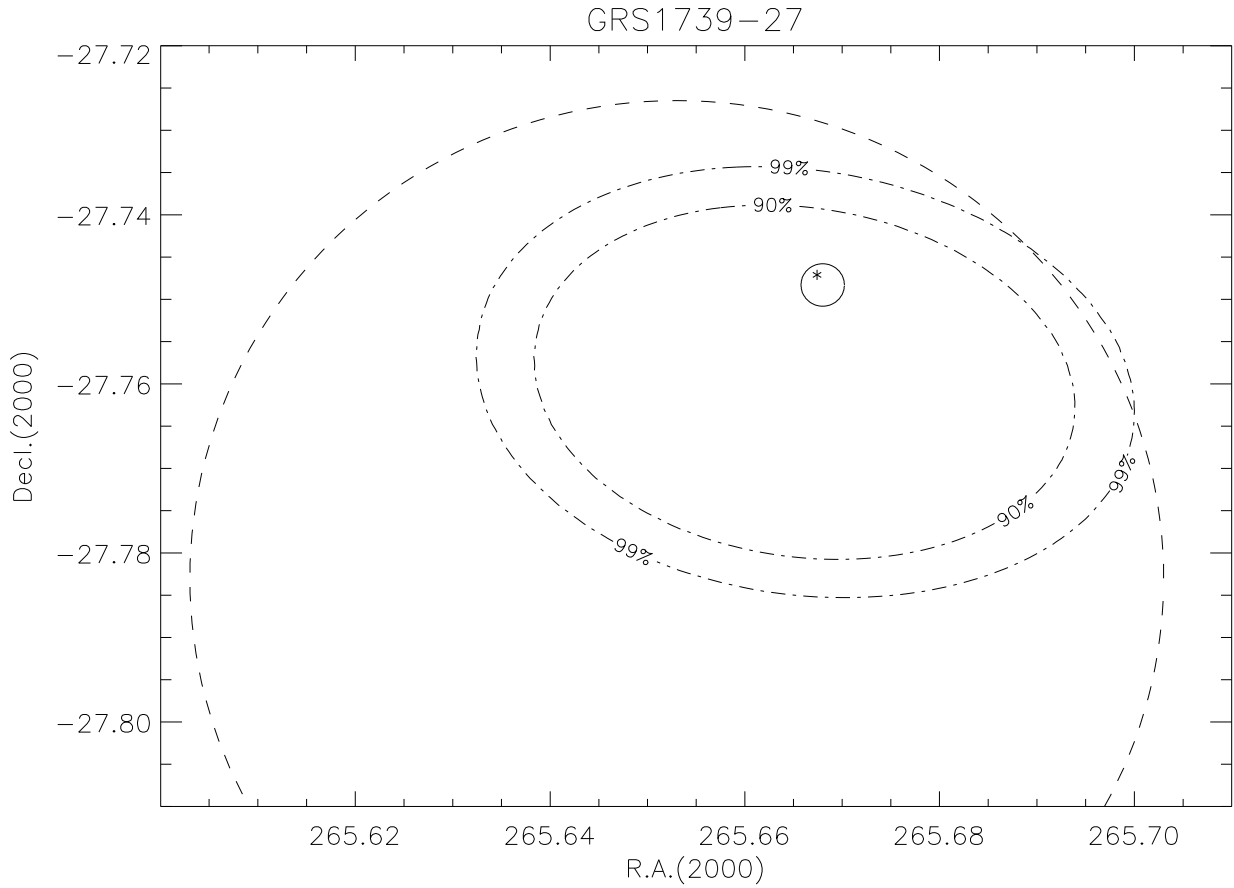} 

\caption{Localization of the X-ray nova GRS 1739-278. The SIGMA error region
(Paul et al. 1996) is indicated by the dashed line; the TTM 90\% and 99\%
confidence regions are indicated by the dash-dotted lines; and the ROSAT
localization (Dennerl and Greiner 1996) is indicated by the solid line. The
asterisk marks the position of the radio and optical objects (Durouchoux et
al. 1996; Mirabel et al. 1996).}
\end{figure}

\begin{figure}
\epsfxsize=17cm
\epsffile{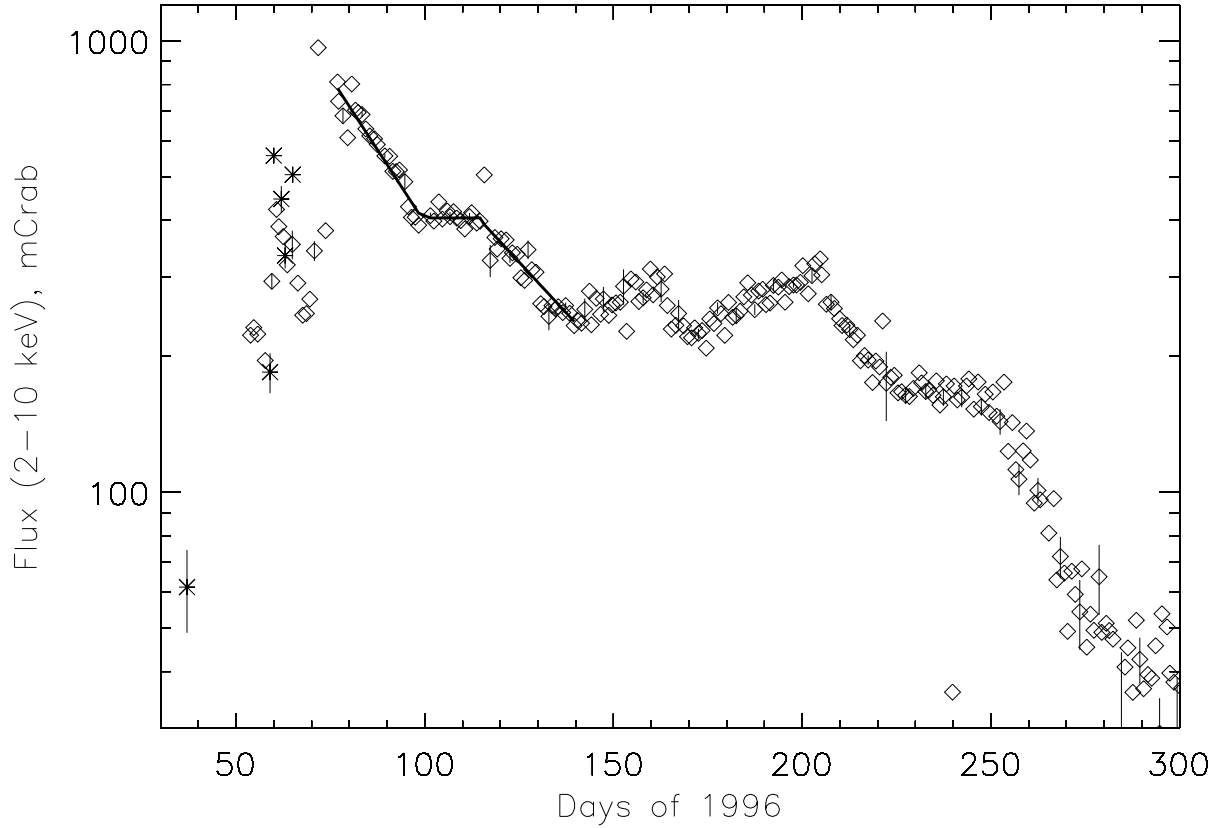} 
\caption{The 2-10-keV light curve of GRS 1739-278, as constructed from the
ASM/RXTE and TTM data. The TTM and ASM data are denoted by the asterisks and
diamonds, respectively. The measurement errors of the flux are given for all
TTM points, and the typical errors are given for some ASM points. The light
curve after the maximum is fitted by a quasi-exponential law (solid line)
with a characteristic time of 34 days (the first 30 days after the maximum),
a constant (days 30-43 after the maximum), and a quasi-exponential law with
a characteristic time of 48 days (days 44-68 after the maximum). }  
     
\end{figure}

\begin{figure}
\epsfxsize=17cm
\epsffile{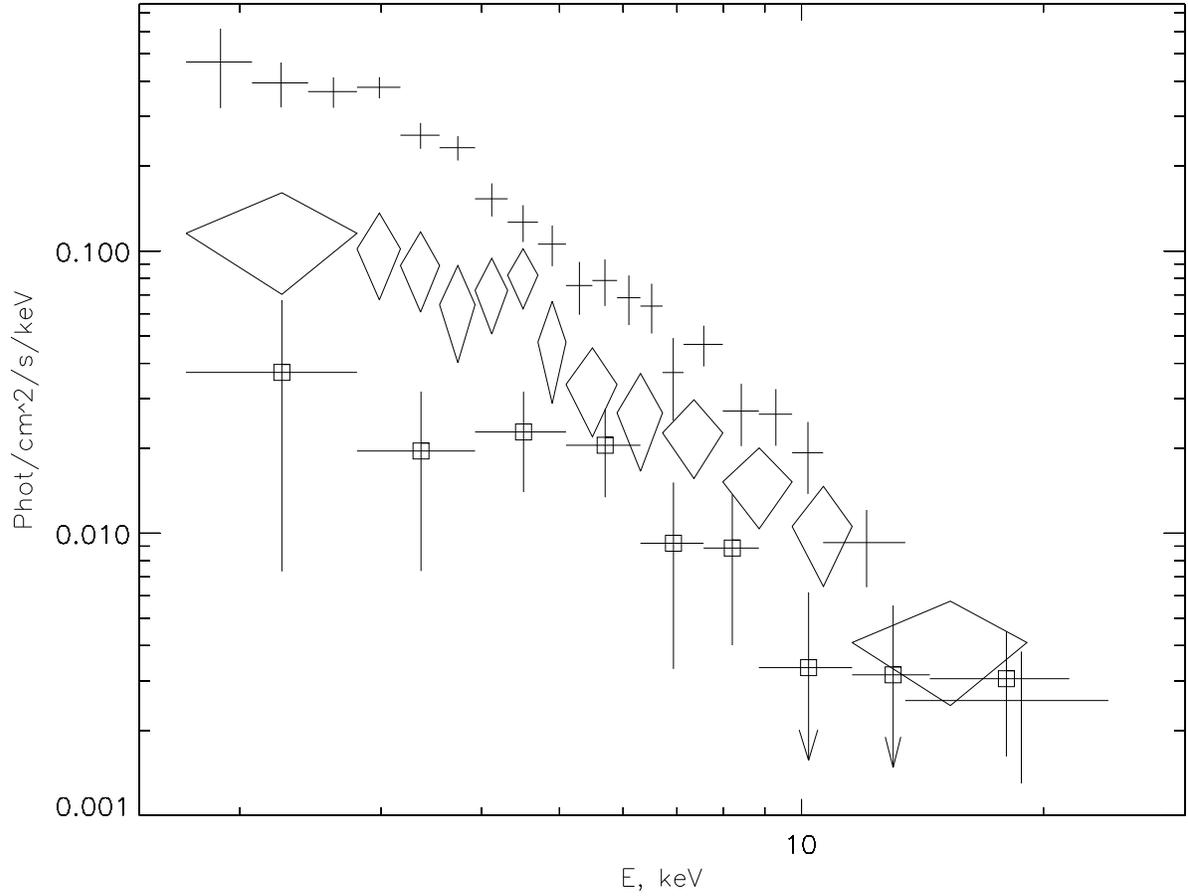} 
\caption{The TTM spectra of GRS 1739-278 on February 6-7 (crosses with
squares), February 28 (diamonds), and March 5 (crosses), 1996. The leading
growth of the soft part of the spectrum is clearly seen. }
 
\end{figure}

\begin{figure}
\epsfxsize=17cm
\epsffile{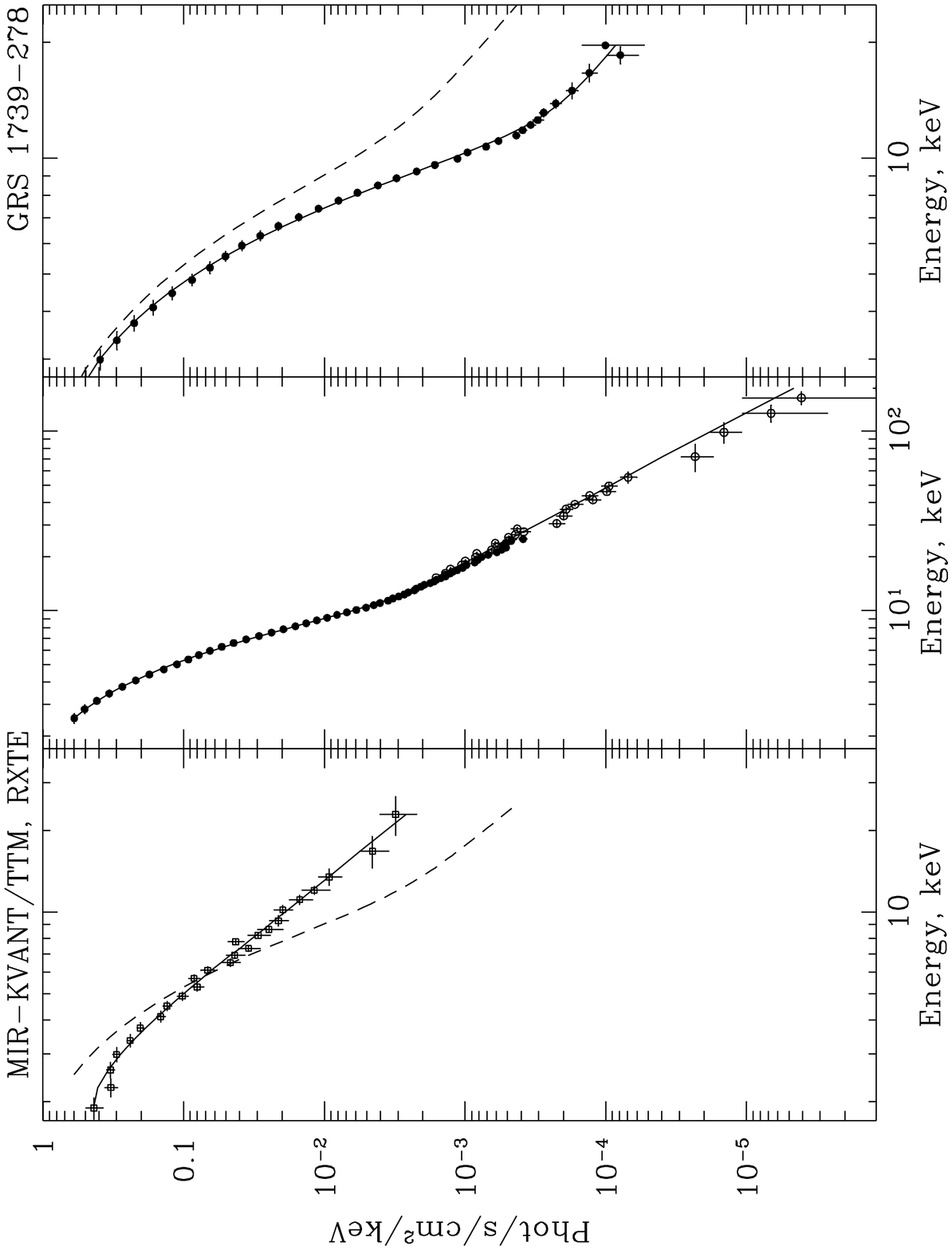} 
\caption{The spectra of GRS 1739-278 during its outburst in 1996: TTM on
March 1-5 (a); RXTE on March 31 (b); the open and filled circles are for
HEXTE and PCA (the relative normalization of the spectra was chosen by using
the PCA data); PCA/RXTE on May 12 (c). The spectra are fitted by a power law
(solid line) with absorption for TTM (see Table 3) and by a two-component
model for RXTE (see Table 4). The fits to the RXTE data on March 31, 1996
are also shown (a and b) for comparison (dashed lines). }      
\end{figure}

\begin{figure}
\epsfxsize=17cm
\epsffile{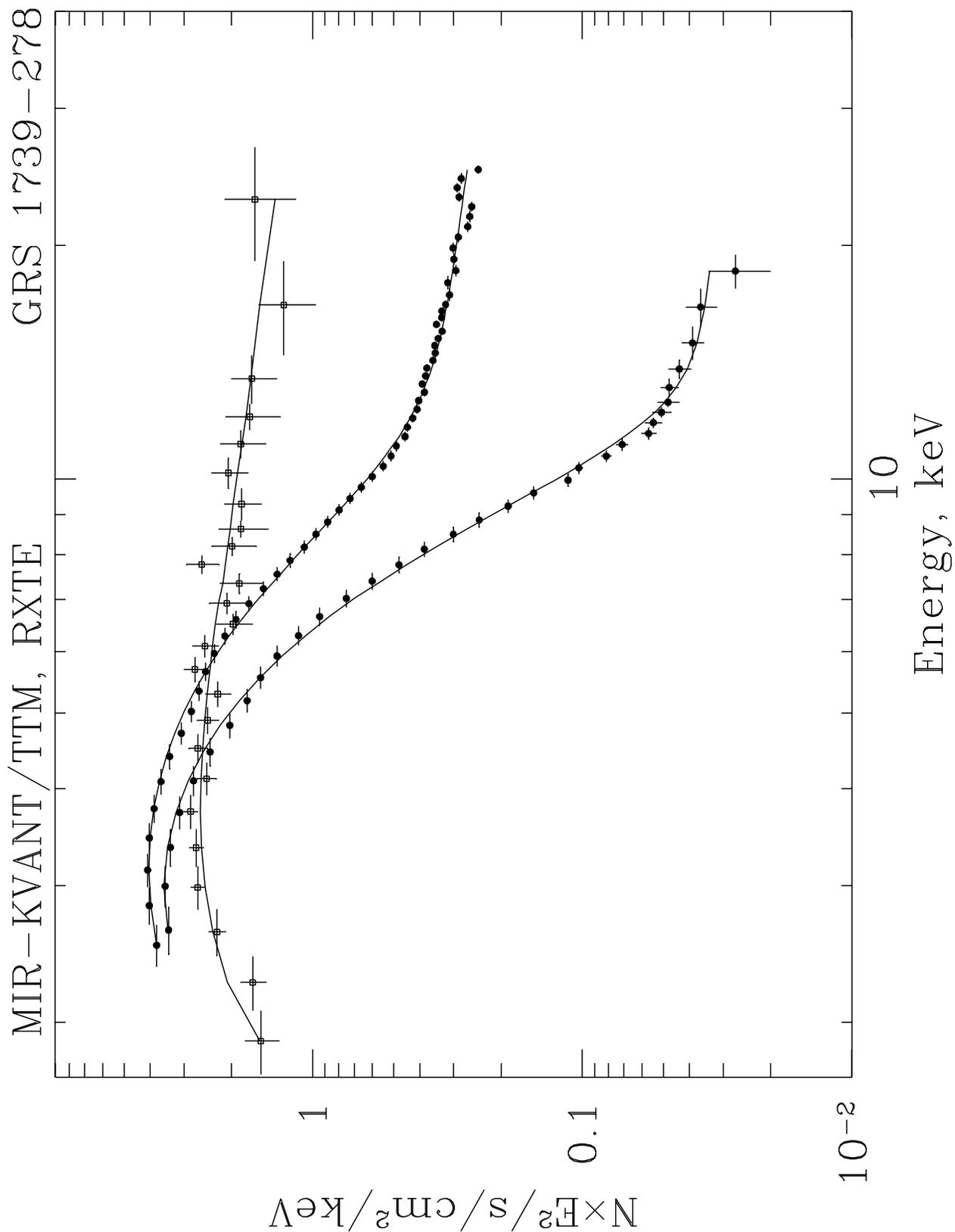} 
\caption{The same spectra of GRS 1739-278 as in Fig. 5, but multiplied by the
energy squared, which clearly shows the relative fraction of energy release
in different bands (only the PCA data are given for the spectrum of March
31).}
\end{figure}

\begin{figure}
\epsfxsize=17cm
\epsffile{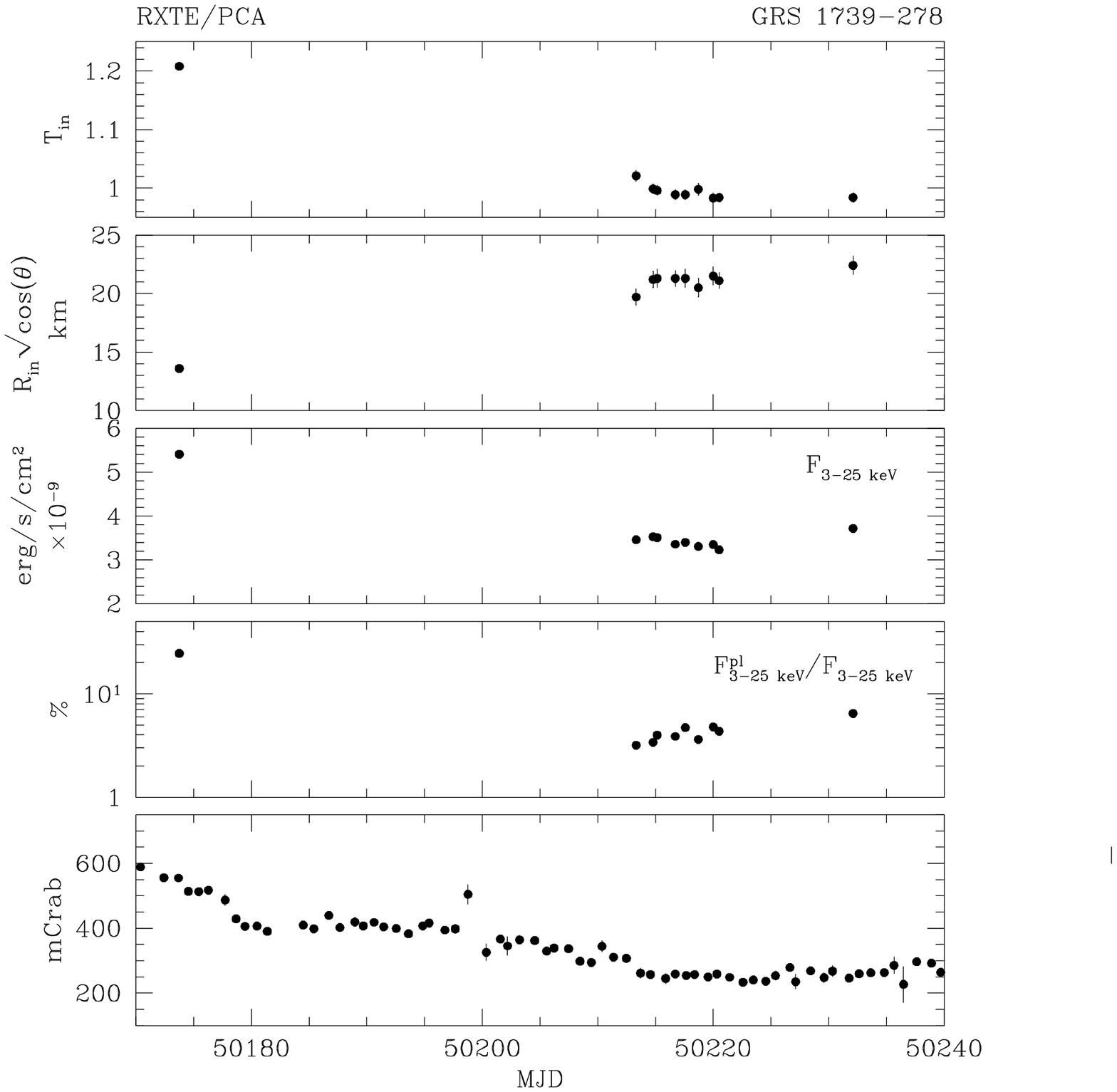} 
\caption{Evolution of the spectrum of GRS 1739-278. In the five panels, the
following fitting parameters for the RXTE spectra are plotted against time:
(a) the model temperature at the inner boundary of the blackbody zone in the
accretion disk (keV); (b) the inner radius of the blackbody disk region
(km); (c) the energy flux from the source in the 3-25-keV band (erg s--1
cm--2); (d) the ratio of the luminosity of the power-law component to the
total luminosity of the source (in percents); and (e) the flux from the
source, as determined from the ASM/RXTE data (in mCrab).  }     
\end{figure}

\begin{figure}
\epsfxsize=17cm

\epsffile{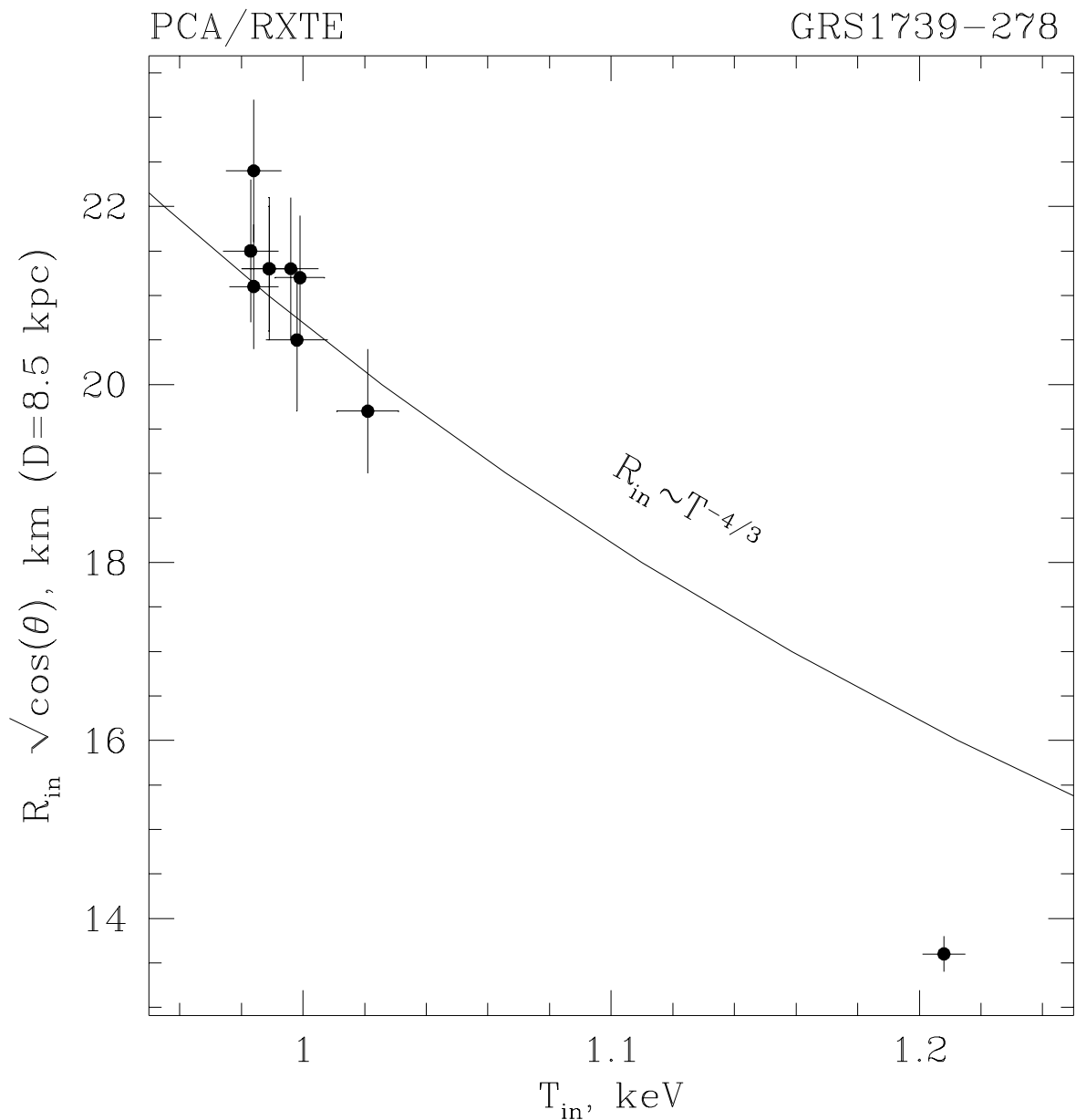} 
\caption{The inner radius of the blackbody disk region versus the model
temperature at the inner boundary of the blackbody zone in the disk. The
solid line represents a theoretical relation between the parameters in the
model of a blackbody accretion disk at a constant accretion rate (Shakura
and Sunyaev 1973). Allowance for the changes in the accretion rate results
in an even greater deviation of the points from the curve. }
\end{figure}

\end{document}